\documentclass[12pt]{article}
\usepackage{hyperref}
\usepackage{lscape}
\usepackage{amsmath}
\usepackage{amssymb}

\renewcommand{\title}[1]{%
    \bigskip%
    \begin{center}%
    \Large\bf #1%
    \end{center}%
    \vskip .2in}

\renewcommand{\author}[1]{%
    {\begin{center}
    #1
    \end{center}}}
\newcommand{\address}[1]{\vspace{-1.7em}\vspace{0pt}
    {\begin{center}
    \it #1
    \end{center}}}

\begin{document}


\title{On the subtelties of nonrelativistic reduction and applications}

\author
{
Rabin Banerjee  $\,^{\rm a,b}$,
Pradip Mukherjee $\,^{\rm c,d}$}
\address{$^{\rm a}$S. N. Bose National Centre 
for Basic Sciences, JD Block, Sector III, Salt Lake City, Kolkata -700 098, India }

\address{$^{\rm c}$Department of Physics, Barasat Government College,\\10, KNC Road, Barasat, Kolkata 700124, India.

 }

\address{$^{\rm b}$\tt rabin@bose.res.in}
\address{$^{\rm d}$\tt mukhpradip@gmail.com}

\begin{abstract}

Various subtelties and problems associated with nonrelativistic (NR) reduction of a scalar field theory to the Schroedinger theory are discussed. Contrary to the usual approaches that discuss the mapping among the equations of motion or the actions, we highlight the mapping among the space time generators. Using a null reduction we show the embedding of the conformal generators of the Schroedinger theory to that of a complex scalar theory in one higher dimension. As applications we reveal the conformal symmetry in hydrodynamics and the obtention of NR diffeomorphism symmetry from the relativistic one. A geometrical connection based on Horava-Lifshitz  and Newton-Cartan spacetime is discussed. 
\end{abstract}

\section{Introduction}
Characterising the nonrelativistic (NR) limit of relativistic quantum field theories is an important issue that is relevant in many contexts, especially when the low energy effective description of a system is required. It has found applications in studying NR diffeomorphism invariance \cite{Andreev:2013qsa}-\cite{Banerjee:2016bbm}
, in fluid dynamics \cite{Horvathy:2009kz,Kaminski:2013gca}, in identifying a plausible dark matter candidate within realistic models of particle physics \cite{Braaten:2016dlp,Namjoo:2017nia} and a host of other phenomena \cite{Leiva:2003kd}-\cite{Ichinose}. Relativistic theories and their equations may be systematically constructed by exploiting the relevant Lorentz symmetry. The obtention of space time generators also follows a well defined track. Such a systematic structure is lacking for NR theories due to the presence of a universal time, contrary to the relativistic case. This has been manifested in no uncertain terms in recent discussions on NR diffeomorphism symmetry and its myriad applications \cite{Andreev:2013qsa}-\cite{Banerjee:2016bbm}. Recourse has been taken to obtain and better understand the results by implementing the NR limit of relativistic expressions. But problems and confusions have persisted, some of which were highlighted in \cite{Banerjee:2016bbm}.

There are different ways of taking the NR limit but this limit may not be unique. In other words the same relativistic theory may have distinct NR limits. It therefore becomes important to identify a specific approach, if possible, which is more suitable, vis-a-vis the other approaches.

Here we address this issue in some details and provide a definitive prescription of taking the NR limit, that emerges naturally from our analysis. It is done for the specific reduction of a relativistic scalar field theory to the Schroedinger theory. Based on this reduction we provide a few examples, both in flat and curved spaces.

Among the various approaches, two are usually favoured in discussing the NR reduction of relativistic theories. These are (i) taking the $c\rightarrow \infty$ limit and (ii) using a light cone basis in one lower dimension. Generally, the NR reduction is analysed at an algebraic level. For example, the Poincare generators are reduced to the Galilean generators by taking the $c\rightarrow\infty$ limit. This is the Inonu-Wigner group contraction where the number of generators in the two (relativistic and NR) cases remains the same. In the light cone analysis, on the other hand, the NR sector is embedded in the relativistic example in one higher dimension so that there is a mismatch in the number of generators. A familiar example is the embedding of the (NR) Schr\"odinger algebra in the (relativistic) conformal algebra in one higher dimension that is explicitly demonstrated using light cone coordinates \cite{Son:2008ye}.

Leaving aside the purely algebraic reductions, investigations have also been performed involving field theories. Usually the NR equations of motion are obtained from the relativistic equations using some prescription. In some cases this equivalence has also been extended to the respective actions. However, this is not enough. It is definitely desirable, if not essential, to show this mapping among the complete set of spacetime generators. A major thrust of the present paper is in this direction.

While the mapping of generators has been accomplished at the algebraic level, using either the Inonu-Wigner group contraction or null reduction, the corresponding situation using field variables has not been explored. As we show there are several pitfalls and subtleties involved here. We discuss these and provide a possible resolution. Taking the relativistic scalar theory as the starting point, we discuss different approaches to consider its NR reduction to the Schroedinger theory. A specific approach is singled out. We find that there is an exact mapping of the action of the massless complex scalar theory in lightcone variables to the Schroedinger theory in one lower dimension. In these variables the complete set of relativistic conformal generators maps to the conformal generators of the Schroedinger theory. As an application, using these generators we discuss the Schroedinger symmetry in hydrodynamics. A more nontrivial application is given in the context of NR diffeomorphism symmetry. The relativistic origin of this symmetry is revealed in the light  cone formulation. Finally, a connection with the projectable version of Horava-Lifshitz geometry and Newton-Cartan geometry is established.

In section 2 we review the quantum mechanical algebraic reduction of the generators. Section 3 and 4 elaborate the NR reduction of a relativistic real scalar theory and a complex scalar theory, respectively in equal time coordinates. There are problems in both cases. Section 5 treats the reduction in a light cone formulation. All results in this case are consistently derived. The generators of the Schroedinger symmetry are obtained from the reduction of conformal generators of a massless complex scalar theory in one higher dimension. This is an exact analogue of the algebraic reduction discussed in Section 2. Sections 6 and 7 contain applications to hydrodynamics and NR diffeomorphism invariance. Specifically, NR spatial diffeomorphism symmetry is obtained from relativistic diffeomorphism symmetry. By including temporal diffeomorphism also, a connection with geometry is shown in Section 8. A consistent NR reduction is possible only by dropping the boost symmetry. In that case the geometry of the projectable Horava-Lifshitz gravity \cite{Horava:2009uw} is reproduced. In Section 9 we show a consistency check of our analysis that is based on Newton-Cartan geometry. Finally our concluding remarks are given in Section 10.
\section{Nonrelativistic reduction at algebraic level}
The conformal algebra is given by,
\begin{align}
 [M_{\mu\nu},  M_{\alpha\beta}]&= -i(\eta_{\nu\alpha} M_{\mu\beta}+\eta_{\mu\beta}M_{\nu\alpha}-\eta_{\mu\alpha}M_{\nu\beta}
  - \eta_{\nu\beta} M_{\mu\alpha})\notag\\
  [P_{\alpha}, M_{\mu\nu}] &= i (\eta_{\nu\alpha} P_{\mu}-\eta_{\mu\alpha} P_{\nu}),~~ [K_{\mu}, M_{\nu\rho}] =i(\eta_{\mu\rho}K_{\nu}-\eta_{\mu\nu}K_{\rho}) \notag\\ [K_{\mu}, P_{\nu}] =& 2i(\eta_{\mu\nu} D + M_{\mu\nu}),~~ [D, P_{\mu}]=iP_{\mu},~~ [D, K_{\mu}] =-i K_{\mu}
\label{ca}
\end{align}
with other commutators vanishing. A particular representation is defined by,
\begin{align}
P_{\mu} &=-i\partial_{\mu}\notag\\
M_{\mu\nu} &= -i(x_{\mu}\partial_{\nu}-x_{\nu}\partial_{\mu})=x_{\mu}P_{\nu}-x_{\nu}P_{\mu}\notag\\
D &=-i x_{\mu}\partial^{\mu}=x_{\mu}P^{\mu}\notag\\ K_{\mu} &=i(x^2\partial_{\mu}-2x_{\mu}x_{\nu}\partial^{\nu})
=-x^2P_{\mu}+2x_{\mu}x_{\nu}P^{\nu}
\label{B}
\end{align}
where $M_{\mu\nu}$, $P_{\mu}$, $D$ and $K_{\mu}$ generate, respectively, Lorentz transformations, translations, scaling (or dilatations) and special conformal transformations.

The NR reduction is done by inserting `$c$' and taking the limit $c\rightarrow\infty$ appropriately. We show this explicitly for the boosts. Recalling $x^0=-x_0=ct$ we obtain,
\begin{equation}
M_{0i}=-i\left(-ct\partial_i-\frac{x_i}{c}\partial_{t}\right)
\end{equation}
from which the NR boost follows as,
\begin{equation}
B_i=\lim_{c\rightarrow\infty} \frac{1}{c}M_{0i}=it\partial_{i}
\end{equation}
Likewise, the other generators are obtained as,
\begin{align}
J_{ij} &=-i(x_i\partial_j-x_j\partial_i),\notag\\P_i &=-i\partial_i,\notag\\H &=\lim_{c\rightarrow\infty}(cP_0)=-i\partial_t\notag\\D &=-i(t\partial_t+x_i\partial_i),\notag\\K &=\lim_{c\rightarrow\infty}(\frac{K_0}{c})=i(t^2\partial_t+2tx_i\partial^i),\notag\\K_i &=\lim_{c\rightarrow\infty}\left(\frac{1}{c^2}K_i\right) =-it^2\partial_i\label{5}
\end{align}
The above set generates the Galilean conformal algebra while the Galilean subalgebra is given by $\left\lbrace H, P_i, B_i, J_{ij}\right\rbrace$. The nontrivial commutators involving the conformal generators are given by,
\begin{align}
& [D,\, P_i] = i P_i, \quad [D,\, H] = iH, \quad [D,\, K]=-iK,\notag\\&[D,\, K_i] = -iK_i, \quad [K,\, P_i]=2iB_i, \quad [K,\, H]=-2iD, \quad \notag\\& [K_i,\,H]=-2iB_i,\quad [K_i,\,J_{kl}]=i(\delta_{il}K_k-\delta_{ik}K_l)
\end{align}
while the purely Galilean sector yields the standard algebra involving $\left\lbrace H, P_i, B_i, J_{ij}\right\rbrace$.

Note that the dilatation generator $D$ scales both space and time uniformly. This is a consequence of obtaining the various NR generators from their relativistic counterparts where space and time are treated on an identical footing. In general, however, space and time can  scale differently in NR theories. In particular, for the Schroedinger theory, scaling of time and space differs by a  factor of two. This suggests the existence of an alternative method of reduction.
  
We next consider the reduction using a dimensional descent in light cone variables. In the earlier approach, the number of generators exactly match in the relativistic and NR cases leading to the Galilean conformal generators from the conformal ones. However the Galilean conformal algebra is not the only NR algebra. The Schrodinger algebra, for instance, is another such algebra and it is embedded in the conformal algebra of one higher dimension. This is shown in light cone coordinates. These coordinates are introduced by defining
\begin{equation}
X^{\pm} = \frac{x^0 \pm x^{d+1}}{\sqrt{2}}
\label{Z}
\end{equation} 
and the metric, 
\begin{equation}
g_{+-} = g_{-+} = g^{+-} = g^{-+} = -1; g_{++} = g_{--} = 0
\end{equation}
Now considering the algebra (\ref{ca}) where $\mu = 0,1,\cdots,d+1$ (i.e. $d+2$ spacetime dimensions), we identify the field light cone momentum by,
\begin{equation}
P^+ = \frac{P^0 + P^{d+1}}{\sqrt{2}}
\end{equation}
with the mass operator $M$ in the NR theory. The generators of the Schrodinger algebra in $d$ spatial dimensions are obtained from (\ref{B}) by the identification \cite{Son:2008ye}, \footnote{the tilde denotes NR generators}
\begin{align}
M = -P^+, &~\tilde{H} = -P^-, \tilde{P}^i =- P^i, \tilde{M}^{ij} = M^{ij} \notag\\
\tilde{K}^i &=M^{i +}, \tilde{D} = -D + M^{+-}, \tilde{C} = \frac{K^+}{2}\label{R}
\end{align}
Using these generators in (1) yields the Schroedinger algebra, 
\begin{equation}\label{Schroed-al}
\begin{split}
  & [\tilde{M}^{ij},\, \tilde{M}^{kl}] = i( \delta^{ik} \tilde{M}^{jl} + 
  \delta^{jl} \tilde{M}^{ik} - \delta^{il} \tilde{M}^{jk} - \delta^{jk} \tilde{M}^{il}),\\
  & [\tilde{M}^{ij},\, \tilde{P}^k] = i(\delta^{ik} \tilde{P}^j - \delta^{jk} \tilde{P}^i), \quad
  [\tilde{M}^{ij},\, \tilde{K}^k] = i(\delta^{ik} \tilde{K}^j - \delta^{jk} \tilde{K}^i), \\
  & [\tilde{D},\, \tilde{P}^i] = -i \tilde{P}^i, \quad [\tilde{D},\, \tilde{K}^i] = i \tilde{K}^i, \quad
    [\tilde{P}^i,\, \tilde{K}^j] = -i\delta^{ij}M,\\
  & [\tilde{D},\, \tilde{H}] = -2i\tilde{H}, \quad [\tilde{D},\, \tilde{C}] = 2i\tilde{C},\quad
    [\tilde{H},\, \tilde{C}] = i\tilde{D},\quad [\tilde{P}^i,\, \tilde{C}]=-i\tilde{K}^i.
\end{split}
\end{equation}
Note that the operator $P^+$ commutes with all other generators justifying its identification with the mass operator $M$. It gives the central extension that can be verified from the nontrivial $\tilde P - \tilde K^i$ (translation- boosts) commutator. This commutator is otherwise vanishing in the Galilean conformal algebra and reveals an important difference between the the two types of NR algebra. There is another important difference that is related to scaling properties. As already mentioned, in the Galilean conformal algebra, time and space are scaled uniformly. In the present example, the dilatation generator has the form,
\begin{equation}
\tilde{D} = -D+M^{+-}= i(2 X^+\partial_{+} + x^i\partial_i)
\end{equation}

Since $X^+$ is identified with the light cone time, we see that the scaling properties of time and space differ by a factor of two. It brings out the the difference from the earlier example. Indeed this form of  scaling occurs in the Schroedinger algebra.
This concludes the algebraic reduction of the conformal generators to the Schrodinger generators. We now analyse the corresponding situation for field variables.

\section{Reduction from a free scalar theory}

The complexities of a nonrelativistic reduction in field theory are already contained in a simple example. Let us consider the nonrelativistic reduction of a real scalar theory (Klein-Gordon theory) to the Schroedinger theory. The starting point is the lagrangian,
\begin{equation}\label{relat}
{\cal{L}} = -\frac{1}{2} \left( \eta^{\mu\nu}\partial_\mu\phi\partial_\nu\phi 
       +m^2c^2\phi^2\right)
\end{equation}
which yields the following equation of motion,
\begin{equation}
 \left( \eta^{\mu\nu}\partial_\mu\partial_\nu 
      -m^2c^2\right)\phi=(\Box-m^2c^2)\phi=0\label{eom}
\end{equation}
A change of variable is introduced,
\begin{equation}\label{Psipsi}
\phi = \frac{e^{-imc^2t}}{\sqrt{2mc}}\psi+\frac{e^{imc^2t}}{\sqrt{2mc}}\psi^*
\end{equation}
where $\psi$ is a complex scalar. Inserting (\ref{Psipsi}) in (\ref{eom}) and using $\Box=-\frac{1}{c^2}\partial_t^2+\partial_i^2$, We find,
\begin{equation}
\frac{e^{-imc^2t}}{\sqrt{2mc}}\left[2im\frac{\partial\psi}{\partial t}+\partial_i^2\psi-\frac{1}{c^2}\frac{\partial^2\psi}{\partial t^2}\right]+\frac{e^{imc^2t}}{\sqrt{2mc}}\left[-2im\frac{\partial\psi^*}{\partial t}+\partial_i^2\psi^*-\frac{1}{c^2}\frac{\partial^2\psi^*}{\partial t^2}\right]=0\label{eem1}
\end{equation}
Taking the $c\rightarrow \infty$ limit and interpreting $\psi, \psi^*$ as independent fields yields the pair of equations,
\begin{align}
i\frac{\partial\psi}{\partial t}+\frac{1}{2m}\partial_i^2\psi&=0\notag\\-i\frac{\partial\psi^*}{\partial t}+\frac{1}{2m}\partial_i^2\psi^*&=0\label{eem}
\end{align}
These are the usual equations of motion for the Schroedinger theory,
\begin{eqnarray}
{\cal{L}} &=&  -\frac{1}{2}\left[i\psi\overleftrightarrow \partial_t\psi^*
   +\frac{1}{m}\partial_i\psi \partial_i\psi^*\right]
\label{globalaction2} 
\end{eqnarray} 
where,
\begin{equation}
\psi\overleftrightarrow\partial_t\psi^*=\psi(\partial_t\psi^*)-(\partial_t\psi)\psi^*
\end{equation}

The interpretation of $\psi$ and $\psi^*$ as independent fields in obtaining (\ref{eem}) from (\ref{eem1}) needs some elaboration. In this context
note that (\ref{eem}) is also obtained from (\ref{globalaction2}) by independent variations of $\psi$ and $\psi^*$. Although $\psi^*$ is the complex conjugate of $\psi$, this is justified since expressing,
\begin{equation}
\psi=\psi_{R}+i\psi_{I}, \psi^*=\psi_R-i\psi_I\label{8}
\end{equation}
in (\ref{globalaction2}), yields the following equations,
\begin{equation}
\partial_t\psi_R+\frac{1}{2m}\partial_i^2\psi_I=\partial_t\psi_I+\frac{1}{2m}\partial_i^2\psi_R=0\label{9}
\end{equation}
where $\psi_R$ and $\psi_I$ are regarded as independent fields. It is now simple to check that (\ref{eem}) is consistent with (\ref{8}) and (\ref{9}).

Although the equations of motion for the NR theory are obtained from the relativistic theory, the same mapping (\ref{Psipsi}) fails to yield the Schroedinger theory (\ref{globalaction2}) from (\ref{relat}). Direct insertion of (\ref{Psipsi}) in (\ref{relat}) yields,
\begin{align}
S&=-\frac{1}{2}\int d^3 x (cdt) \left[\frac{i}{c}\psi\overleftrightarrow\partial_t\psi^*+\frac{1}{cm}\partial_i\psi\partial_i\psi^*-\frac{\dot{\psi}\dot{\psi}^*}{mc^3}\right.\notag\\&\left.+\frac{i}{4c}\frac{d}{dt}(e^{-2imc^2t}\psi^2-e^{2imc^2t}\psi^{*2})+\frac{1}{2mc}(e^{-2imc^2t}(\partial_i\psi)^2+e^{2imc^2t}(\partial_i\psi^*)^2)\right.\notag\\&\left.-\frac{1}{2mc^3}(e^{-2imc^2t}\dot{\psi}^2+e^{2imc^2t}\dot{\psi}^{*2})\right]
\end{align} 
On taking the $c\rightarrow\infty$ limit, the first pair of terms yields the desired result and the next drops out. The next piece may also be dropped as it is a total time derivative. But the next terms remain. Only if we use an approximation where rapidly oscillating terms  (proportional to $e^{{\pm}2imc^2t}$) may be ignored, do we reproduce the expected result. This approximation, however, may not be always valid \cite{Namjoo:2017nia}. Thus although the reduction from the relativistic to nonrelativistic holds for the equation of motion, there is a problem at the level of lagrangian or the action. The origin of this problem may lie in the proposed mapping of a real scalar theory to a complex scalar theory. Indeed, the complex scalar theory given by the Schroedinger lagrangian has a $U(1)$ gauge invariance whose generator is given by the central charge $P^+$ proportional to the mass. There is no analogue of this in the Lorentz symmetry group of the Klein-Gordon theory. The wavefunction provides a projective representation of the Galilei group and hence must be complex to yield the additional $U(1)$ invariance.This induces us to look into the possibility of mapping a (relativistic) complex scalar theory, which has the requisite $U(1)$ invariance,  to the Schroedinger theory, which is the topic of the next section. 


\section{Reduction of complex scalar theory}
 Let us consider a theory of free complex scalars,
\begin{equation}
S=-\int d^4 x (\eta^{\mu\nu}\partial_{\mu}\phi^*\partial_{\nu}\phi+m^2 c^2 \phi^*\phi)\label{action}
\end{equation}
To perform a NR transition, the change of variables,
\begin{equation}
\phi=e^{-imcx^0}\frac{\psi}{\sqrt{2mc}}=e^{-imc^2t}\frac{\psi}{\sqrt{2mc}}\label{nr}
\end{equation}
is introduced. Inserting in (\ref{action}) we find,
\begin{equation}
S=\int d^3x dt\left[\frac{i}{2}\psi^*\overleftrightarrow{\partial_t}\psi-\frac{\partial_i\psi^*\partial_i\psi}{2m}+\mathcal{O}\left(\frac{1}{c^2}\right)\right]\label{psi}
\end{equation}

Taking the $c\rightarrow\infty$ limit in (\ref{psi}) yields the NR (free) Schroedinger theory. This is a well known result. Moreover, since the mapping is shown for the respective lagrangians, the equations of motion will also map accordingly.

Let us now probe this mapping further. The stress tensor following from (\ref{action}) is given by,
\begin{equation}
T_{\mu\nu}=-(\partial_{\mu}\phi^*\partial_{\nu}\phi+\partial_{\mu}\phi
\partial_{\nu}\phi^*)+\eta_{\mu\nu}(\partial^{\sigma}\phi^*\partial_{\sigma}\phi+m^2c^2\phi^*\phi)\label{stress}
\end{equation}
which is obtained either by using Noether's definition, 
\begin{equation}
T_{\mu\nu} = \frac{\partial \mathcal{L}}{\partial (\partial^{\mu} \phi)} \partial_{\nu} \phi + \frac{\partial \mathcal{L}}{\partial (\partial^{\mu} \phi^{*})} \partial_{\nu} \phi^{*}- \mathcal{L} \eta_{\mu \nu}
\label{Noeth}
\end{equation}
or, by varying with respect to the metric before setting it flat,
\begin{equation}
T_{\mu\nu} =- \frac{2}{\sqrt{-g}} \frac{\delta{S}}{\delta{g^{\mu \nu}}} = 2 \frac{\delta \mathcal{L}}{\delta g^{\mu \nu}} - \mathcal{L} g_{\mu \nu}
\label{st}
\end{equation}
The hamiltonian density follows from (\ref{stress}),
\begin{equation}
P_0=-T_{00}=\frac{1}{c^2}\dot{\phi}^*\dot{\phi}+\partial_i\phi^*\partial_i\phi+m^2c^2\phi^*\phi
\end{equation}
Using the map (\ref{nr}) we find,
\begin{equation}
P_0=mc\psi^*\psi+\frac{\partial_i\psi^*\partial_i\psi}{2mc}+\frac{i}{2c}(\psi^*\dot{\psi}-\dot{\psi}^*\psi)+\frac{\dot{\psi}^*\dot{\psi}}{2mc^3}
\end{equation}
The time derivatives are eliminated by using the equations of motion for $\psi, \psi^*$,
\begin{equation}
\dot{\psi}=\frac{i}{2m}\partial_i^2 \psi \, , \dot{\psi}^*= - \frac{i}{2m}\partial_i^2 \psi^*
\end{equation} 
so that
\begin{equation}
c P_0 = m c^2  \psi \psi^* + \frac{1}{2m} \partial_i\psi \partial_i\psi^* - \frac{1}{4m}\left(\psi\partial_i^2 \psi^* + \psi^*\partial_i^2 \psi\right)
\label{ed}
\end{equation} 
and a term of $\mathcal{O}(\frac{1}{c^2})$ has been dropped. Identifying $cP_0$ with the NR energy density (see the analogous relation for the algebraic representation (\ref{5})) we observe the first term on the r.h.s. of  (\ref{ed}) may be interpreted as the zero point (rest) energy density followed by the expected terms. However, this does not hold since the expression obtained directly from (\ref{psi}) does not contain the last bracketed pair of terms in (\ref{ed}). This pair does not vanish even if the integrated version is considered.

\section{Light cone reduction}
\bigskip
 The failure of the complex scalar theory to yield a consistent NR reduction to the Schroedinger theory might be related to the fact that, whereas the former is a second order theory, the latter is first order. A nontrivial change in the symplectic structure is needed. This is most easily done by using the light cone coordinates. In this formulation the relativistic scalar theory has a manifestly `nonrelativistic" look from the outset.
 
The starting point is a free theory of massless complex scalars which is conformal invariant.
\begin{equation}
\mathcal{L} = -\frac{1}{2}\partial_{\mu}\phi^* \partial^{\mu} \phi  \qquad (\mu= 0,1,2,3)
\label{L}
\end{equation}
Introducing the null coordinates (\ref{Z}) with $d=2$ one obtains
\begin{equation}
\mathcal{L} = \frac{1}{2} \left(\partial_{+}\phi^* \partial_- \phi + \partial_{-}\phi^* \partial_+ \phi -\partial_m \phi^* \partial_m \phi\right) \quad (m=1,2)
\label{as} 
\end{equation}
where the variable $X^+$ is identified with the time coordinate. The important point is that the above lagrangian is first order in time derivative which is what we wanted. Now the mass $M$ in the Schrodinger algebra was mapped to light cone momentum $P^+ \sim \frac{\partial}{\partial X_+} = \frac{\partial}{\partial X^-}$. Compactifying $X^{-}$ would naturally yield a discrete spectrum for the mass which is expected in a NR theory. This suggests the following change of variables,
\begin{align}
\phi(X^+,X^-,x^1,x^2) &= e^{-imX^-} \psi(X^+,x^1,x^2)\notag\\
\phi^*(X^+,X^-,x^1,x^2) &= e^{imX^-} \psi^*(X^+,x^1,x^2)
\label{MAP}
\end{align}
Inserting this in the lagrangian (\ref{as}) yields
\begin{equation}
\mathcal{L} = m \left(\frac{i}{2}\psi^*\overleftrightarrow{\partial_+}\psi
-\frac{\partial_m\psi^*\partial_m\psi}{2m}\right)
\label{X}
\end{equation}
which, modulo a trivial normalization, reproduces the NR Schrodinger theory with mass $m$ in one lower space dimension. Observe that the spatial coordinate $X^-$ has been compactified and disappears from (\ref{X}).  Contrary to the obtention of (\ref{psi}), this result has been derived exactly without the imposition of any limits. 

The stress tensor following from (\ref{L}) is given by,
\begin{equation}
T_{\mu\nu}=-\frac{1}{2}(\partial_{\mu}\phi^*\partial_{\nu}\phi+ \partial_{\mu}\phi
\partial_{\nu}\phi^*)+\frac{1}{2} \eta_{\mu\nu}(\partial^{\sigma}\phi^*\partial_{\sigma}\phi)
\label{lebel}
\end{equation}
The various conformal generators will be obtained from (\ref{lebel}). For instance, the four momenta is given by,
\begin{equation}
P^{\mu} = \int T^{+\mu}
\label{sht}
\end{equation}
from which the Hamiltonian has the form
\begin{equation}
H=-P^{-} = -\int T^{+ -} = -\int T_{-+} = \frac{1}{2} \int \partial_m \phi^* \partial_m \phi  
\label{crp}
\end{equation}
This is the expected structure since, in null coordinates, the lagrangian (\ref{L}) has the first order form
\begin{equation}
\mathcal{L} = \frac{1}{2} (\partial_{+}\phi^* \partial_- \phi + \partial_{-}\phi^* \partial_+ \phi ) -\frac{1}{2}(\partial_m \phi^* \partial_m \phi)
\label{assh} 
\end{equation}
which yields the hamiltonian (\ref{crp}).

On using the map (\ref{MAP}) the hamiltonian (\ref{crp}) takes the form,
\begin{equation}
\tilde{H}=-P^{-}=\frac{1}{2} \int \partial_m \phi^* \partial_m \phi = \frac{1}{2} \int \partial_m \psi^* \partial_m \psi
\label{pss}
\end{equation}
which reproduces the hamiltonian of the NR Schrodinger theory trivially read-off from (\ref{X}). Thus the algebraic relation $\tilde{H} = -P^-$ in (\ref{R}) is satisfied at the operator level.

From (\ref{sht}) the spatial translation for the relativistic theory is defined by
\begin{equation}
P_{i} = \int T^{+}_{i} = -\int T_{-i} = \frac{1}{2} \int (\partial_{-}\phi^* \partial_i \phi + \partial_{-}\phi \partial_i \phi^* )
\label{crp2}
\end{equation}
Exploiting the map (\ref{MAP}), the corresponding expression for the Schrodinger theory becomes,
\begin{equation}
\tilde{P}_{i}=-{P}_{i}=-\frac{im}{2}\int(\psi\partial_i\psi^*-\psi^*\partial_i\psi)\label{32}
\end{equation}
which is the expected result. This may be easily verified from the expression obtained from (\ref{X}), defined by Noether's prescription,
\begin{align}
\tilde{P}_{i}&=-\int \tilde{T}_{-i}=-\int \left(\frac{\partial \mathcal{L}}{\partial (\partial^{-} \psi)} \partial_{i} \psi + \frac{\partial \mathcal{L}}{\partial (\partial^{-} \psi^{*})} \partial_{i} \psi^{*}\right) \notag\\&=-\frac{im}{2}\int(\psi\partial_i\psi^*-\psi^*\partial_i\psi)
\label{321}
\end{align}
It is simple to show, using the basic algebra following from the symplectic structure of (\ref{X}),
\begin{equation}
[\psi(x), \psi^*(y)]=\frac{1}{m}\delta(x-y)
\end{equation}
that $\tilde{P}_i$ generates the expected translations,
\begin{equation}
[\psi, \tilde{P}_i]=i\partial_i\psi\label{ft}
\end{equation}
Likewise, the Hamiltonian (\ref{pss}) generates the equation of motion,
\begin{equation}
i\partial_+\psi=[\psi, \tilde{H}]=-\frac{1}{2m}\partial_m^2\psi
\end{equation}
which reproduces the Euler-Lagrange equation following from (\ref{X}).
Finally,
\begin{equation}
P^{+}=\int T^{++}=\int T_{--}=-\int \partial_{-}\phi^*\partial_{-}\phi
\end{equation}
Exploiting the map (\ref{MAP}) we find,
\begin{equation}
P^{+}=-m^2\int \psi^*\psi\label{37}
\end{equation}
 Thus the operator $P^{+}$ gets identified with the NR mass term. This completes the proof of the first three relations in (\ref{R}).

Next, consider the angular momentum operator,
\begin{equation}
M_{ij}=-\int (x_i T_{-j}-x_{j}T
_{-i})
\end{equation}
Using the expression (\ref{lebel}), we obtain,
\begin{equation}
M_{ij}=\frac{1}{2}\left[x_i(\partial_{-}\phi^*\partial_{j}\phi+
\partial_{-}\phi\partial_{j}\phi^*)-x_j(\partial_{-}\phi^*\partial_{i}\phi+
\partial_{-}\phi\partial_{i}\phi^*)\right]
\end{equation}
Exploiting the map (\ref{MAP}) and the identification (\ref{R}), the corresponding expression for the Schroedinger theory is found,
\begin{equation}
\tilde{M}_{ij}= M_{ij}= -\frac{im}{2}\int[x_i(\psi\partial_j\psi^*-\psi^*\partial_j\psi)
-x_j(\psi\partial_i\psi^*-\psi^*\partial_i\psi)]\label{40}
\end{equation}
This is identical to the result derived directly from (\ref{X}), using Noether's prescription, 
\begin{align}
\tilde{M}_{ij}&=\int x_{i}[(\frac{\partial \mathcal{L}}{\partial (\partial_{+} \psi)} \partial_{j} \psi + \frac{\partial \mathcal{L}}{\partial (\partial_{+} \psi^{*})} \partial_{j} \psi^{*})\notag\\&-x_{j}\left(\frac{\partial \mathcal{L}}{\partial (\partial_{+} \psi)} \partial_{i} \psi + \frac{\partial \mathcal{L}}{\partial (\partial_{+} \psi^{*})} \partial_{i} \psi^{*}\right)]
\end{align}
which completes the demonstration of the fourth relation in (\ref{R}).

The Galilean boosts will be obtained from $M^{i+}$, defined in the relativistic theory, by,
\begin{align}
M^{i+} &=\int (x^i T^{++}-X^{+}T^{+i})=\int x^i T^{++}-X^{+}P^{i}\notag\\ &=-\int x^i\partial_-\phi^*\partial_-\phi-X^+P^i
\end{align}
Observing that the nonrelativistic momentum differs by a sign from its relativistic counterpart, (see (\ref{crp2}), \ref{32}), the expression for the NR boost generator follows as,
\begin{equation}
\tilde{K}^i=M^{i+}=-m^2\int x_i \psi^* \psi+X^+\tilde{P}^i\label{43}
\end{equation}
where the map (\ref{MAP}) has been used.
The dilatation operator for the relativistic theory is given by,
\begin{equation}
D = \int X_{\mu}T^{+\mu}=-\int(X^- T^{++}+X^+ T^{+-}-x^i T^{+i})
\end{equation}
while the operator $M^{+-}$ is given by,
\begin{equation}
M^{+-}=\int(X^+ T^{+-}-X^- T^{++})
\end{equation}
Then the nonrelativistic dilatation generator will be defined as,
\begin{align}
\tilde{D}= &-D+M^{+-}=\int(2X^+ T^{+-}-x^i T^{+i})\notag\\= &-2 X^+\tilde{H}+
\int x^i\tilde{T}^{+i}\label{DD}
\end{align}
where $\tilde{T}^{+i}$ has been defined in (\ref{32}).

The special conformal generator $\tilde{C}$ will be obtained from $K^+$ (see (\ref{R})) where the relativistic generator $K^{\mu}$ is given by (see (\ref{B})),
\begin{equation}
K^{\mu}=-\int (x^2 T^{+\mu}-2x^{\mu}x_{\nu}T^{+\nu})
\end{equation}
so that, after appropriate mappings,
\begin{equation}
\tilde{C}=\frac{K^+}{2}=(X^{+})^2 \tilde{H}+\frac{m^2}{2}\int x_i^2\psi^*\psi
-X^+\int x^i\tilde{T}^{+i}
\end{equation}
Using (\ref{DD}) this form may be simplified further,
\begin{equation}
\tilde{C}=-X^+\tilde{D}-(X^{+})^2 \tilde{H}+\frac{m^2}{2}\int x_i^2\psi^*\psi\label{C}
\end{equation}
It is straightforward to check that the field generators $(\tilde{H}, \tilde{P}_i, \tilde{M}_{ij}, \tilde{K}^i, \tilde{C})$ satisfy the complete conformal algebra (\ref{Schroed-al}). As a specific example, let us take the $\tilde{D}$-$\tilde{P}^i$ bracket,
\begin{equation}
[\tilde{D}, \tilde{P}^i]=[-2X^+\tilde{H}+\int x^j\tilde{T}^{+j}(X), \tilde{P}^i]
\end{equation}
Since $\tilde{P}^i$ generates field translations (see (\ref{ft})), we obtain,
\begin{equation}
[\tilde{D}, \tilde{P}^i]= i\int x^j\partial^i\tilde{T}^{+j}(x)=-i\int(\partial^ix^j)\tilde{T}^{+j}(x)
\end{equation}
obtained on dropping a boundary term. Thus,
\begin{equation}
[\tilde{D}, \tilde{P}^i]=-i\int \tilde{T}^{+i}=-i\tilde{P}^i
\end{equation}
thereby reproducing the desired form. Incidentally the expressions for the field generators obtained here agree with those found earlier \cite{Hagen:1972pd} by using basic properties of the symmetry transformations. 
\section{Schroedinger symmetry in hydrodynamics}
A familiar practical example of the occurence of Schroedinger symmetry is in the context of hydrodynamics. This is not surprising since, by a simple change of variables, it is possible to obtain an action for hydrodynamics from the Schroedinger theory (\ref{X}). Expressing the complex field $\psi$ in terms of its magnitude and phase by,
\begin{equation}
\psi=\sqrt{\rho}e^{i\theta}\label{B1}
\end{equation}
the lagrangian (\ref{X}) yields, on dropping a trivial normalisation (m) factor,
\begin{equation}
{\cal{L}}=-\rho(\partial_+\theta+\frac{1}{2m}(\partial_i\theta)^2)-\frac{(\partial_i\rho)^2}{8m\rho}\label{B2}
\end{equation}
This is the standard lagrangian for an irrotational fluid in Euler variables $(\rho, \theta)$ with a potential $V(\rho)=\frac{(\partial_i\rho)^2}{8m\rho}$, expressed in our light cone system.

The various generators for the theory (\ref{B2}) are now obtained from the Schroedinger theory results, exploiting the map (\ref{B1}).

The translation generator, for instance, is given by,
\begin{equation}
\tilde{P}_{i}=-\frac{i}{2}\int(\psi\overleftrightarrow{\partial}_i\psi^{*})=-\int
\rho\partial_i\theta\label{B3}
\end{equation}
On using (\ref{32}) and (\ref{B1}). The hamiltonian follows from (\ref{pss}),
\begin{align}
\tilde{H}=&\frac{1}{2}\int \partial_i\psi^{*}\partial_i\psi\notag\\&=\frac{1}{2}\int
[\rho(\partial_i\theta)^2+\frac{1}{4\rho}(\partial_i\rho)^2]\label{B4}
\end{align}
while the boosts take the form (\ref{43}),
\begin{equation}
\tilde{K}^i=- \int x^i( \psi^{*} \psi)+x^+\tilde{P}^i=-\int\rho(x^i+X^+\partial^i\theta)\label{B5}
\end{equation}
where (\ref{B3}) also has been used.

The special conformal generator (expansion) is likewise obtained from (\ref{C}) as,
\begin{equation}
\tilde{C}=-X^+\tilde{D}-(X^{+})^2 \tilde{H}+\frac{1}{2}\int \rho x_i^2\label{B6}
\end{equation}
where $\tilde{H}$ is the hamiltonian (\ref{B4}) and the dilatation operator is obtained from (\ref{DD}),
\begin{align}
\tilde{D}= -2 X^+\tilde{H}-\int \rho(x^i\partial^i\theta)\label{B7}
\end{align}
where we have also used (\ref{B3}).

The angular momentum is obtained trivially from (\ref{40}),
\begin{equation}
\tilde{M}_{ij}=-\int\rho[x_i\partial_j\theta-x
_j\partial_i\theta]\label{B8}
\end{equation}
Lastly, the galilean mass is introduced from (\ref{37}) as,
\begin{equation}
M=\int\rho\label{B9}
\end{equation}
The expressions (\ref{B3})-(\ref{B9}) agree with those found by applying geometric arguments and Noether's prescription \cite{Duval:2009vt}.

\section{Curved Background}
Nonrelativistic theories in a curved background have attracted considerable attention due to their various applications. A new approach was developed by us in a series of papers \cite{Banerjee:2014pya}- \cite{Banerjee:2017rch}. Essentially this consisted in gauging the nonrelativistic galilean symmetry which was christened as galilean gauge theory. Here we show that this theory may be obtained by taking a NR limit of a relativistic theory, using the light cone formulation discussed in the previous section.

The lagrangian for a theory of complex scalars in a curved background is written from a simple generalisation of (\ref{L}),
\begin{equation}\label{A1}
  {\cal L} = -\frac{1}{2}g^{\mu\nu}\partial_\mu\phi^*\partial_\nu\phi 
\end{equation}
Expressed in light cone coordinates, it takes the form,
\begin{align}
\cal L=&-\frac{1}{2} [g^{++}\partial_+\phi^*\partial_+\phi+g^{+-}\partial_+\phi^*\partial_-\phi+g^{+i}\partial_+\phi^*\partial_i\phi\notag\\&+g^{-+}\partial_-\phi^*
\partial_+\phi+g^{--}\partial_-\phi^*\partial_-\phi+g^{-i}\partial_-
\phi^*\partial_i\phi+g^{i+}\partial_i\phi^*\partial_+\phi\notag\\&
+g^{i-}\partial_i\phi^*\partial_-\phi+g^{ij}\partial_i\phi^*\partial_j\phi]
\label{A2}
\end{align}
Now making the change of variables (\ref{MAP}), we obtain,
\begin{align}
\cal{ L}=& -\frac{m}{2} [\frac{g^{++}}{m}\partial_+\psi^*\partial_+\psi+ g^{+-} i \psi^* \overleftrightarrow{\partial_+} \psi \notag\\
&+ m g^{--} \psi^* \psi  +  \frac{g^{+i}}{m}\left(\partial_+\psi^*\partial_i\psi + \partial_i\psi^* \partial_+\psi \right)
 + \frac{g^{ij}}{m}\partial_i\psi^*\partial_j\psi \notag\\
&+ i g^{i-} \psi^* \overleftrightarrow{\partial}_i \psi ]
\label{A3}
\end{align}
Let us now recall some results of galilean gauge theory that eventually yield a nonrelativistic diffeomorphism invariant [NRDI] theory \footnote{Here for simplicity, we consider only spatial NRDI. 
} whose action is given by,
\begin{eqnarray}\label{free-L}
  S &=&  \int dt\,d\vec{x}\, \sqrt{\tilde g}\left[\frac{i}{2} \psi^{*} \overleftrightarrow{\partial_t}\psi
  - B_0\psi^{*}\psi 
  - \frac{g^{ij}}{2m}(\partial_i\psi^{*} -  i B_i\psi^{*})(\partial_j \psi + i B_j\psi)\right]\nonumber\\
& & + \int dt\,d\vec{x}\, \sqrt{\tilde g}~ \frac{i}{2}~\Delta^{k} 
 \left[\psi^{*}(\partial_{k}\psi +iB_k \psi) - (\partial_{k}\psi^{*}-iB_k \psi^{*})\psi\right]
\label{A4}
\end{eqnarray}
Note that $\tilde{g} = \text{det} g_{ij}$ where $g_{ij}$ is the spatial metric. Here $B$ and $\Delta$ are new (external) fields introduced on gauging the original flat space theory (\ref{X}). Putting them to zero leads to a flat metric (as we will see soon) and reproduces the flat space theory (24). The above action is invariant under the following infinitesimal transformations \cite{Banerjee:2015rca},
\begin{align}
\delta \psi & = -\xi^k \partial_k \psi \notag\\
\delta B_0 &= -\xi^k \partial_k B_0 - B_k \dot{\xi}^k \notag\\
\delta B_i &= -\xi^k \partial_k B_i - B_k \partial_i \xi^k \notag\\
\delta \Delta_i &= g_{ij} \dot{\xi}^j - \xi^k \partial_k \Delta_i - \Delta_k \partial_i \xi^k \notag\\
\delta g_{ij} &= -\xi^k \partial_k g_{ij} - g_{ik} \partial_j \xi^k - g_{k j} \partial_i \xi^k
\label{A5}
\end{align}
while the diffeomorphism parameter $\xi^i$ corresponding to the shift,
\begin{equation}
x^i \to x^i + \xi^i
\label{A6}
\end{equation}
is space time dependent.

Comparing (\ref{A3}) with the integrand of  (\ref{A4}) yields the following identification,
\begin{align}
g^{++}&= g^{+i} = 0 \, , \qquad  g^{+ -} = -1 \, , \qquad  g^{ij}= g^{ij} \notag\\
g^{--} &= \frac{2B_0}{m} + \frac{B^iB_i}{m^2} + \frac{2 \Delta^i B_i}{m} \,, \quad g^{i-}= -\left(\frac{B^i}{m} + \Delta^i\right)
\label{A7}
\end{align}
which defines the inverse metric, where $B^i = g^{ij} B_j$. The metric is therefore given by,
\begin{equation}
g_{\mu \nu} = 
\begin{pmatrix}
-\frac{2B_0}{m} + \Delta^i \Delta_i & -1 & -(\frac{B_i}{m} + \Delta_i)  \\ 
- 1 & 0 & 0\\
-(\frac{B_i}{m} + \Delta_i)  & 0 & g_{ij}\\
\end{pmatrix}
\label{A8}
\end{equation}
whose determinant is $\tilde{g} = \text{det} g_{ij}$. Observe that this is the determinant of the two by two  matrix corresponding to $x_i$. The dependence on the $X^-$ has dropped out. Also, there is no dependence of $X^-$ on the matter sector in (\ref{A3}). This indicates that in order to achieve a complete mapping of the corresponding actions, the light cone theory has to be defined in one dimension less, so that the dependence of $X^-$ in the metric is dropped,
\begin{equation}
\partial_-g_{\mu\nu}=0
\end{equation}
Then the action corresponding to (\ref{A3}) will be written as,
\begin{equation}
S=\int dX^- dx^1 dx^2  \sqrt{-g} \cal L
\end{equation}
and a concrete mapping of this theory with that of (\ref{A4}) defined in three space time dimensions is possible.

We next show that the metric components given in (\ref{A7},\ref{A8}) satisfy the expected transformation laws, using (\ref{A5}). The covariant transformation law is given by,
\begin{equation}
\delta g_{\mu \nu} = -\xi^{\lambda} \partial_{\lambda} g_{\mu \nu} - g_{\lambda \nu} \partial_{\mu}\xi^{\lambda} - g_{\lambda \mu} \partial_{\nu}\xi^{\lambda}
\label{A9}
\end{equation}
Since we consider spatial diffeomorphisms, we take,
\begin{equation}
\xi^{\mu} = (\xi^+ = 0 , \xi^- = 0, \xi^i )
\label{A10}
\end{equation}
Moreover, since in lightcone variables, the coordinate $X^-$ has been compactified, $\xi^- =0$ and the parameter $\xi^{\mu}$ is a function of $X^+ \, , x^i$ only, $\xi^{\mu} = \xi^{\mu} (X^+,x^i)$.

It is then straightforward to show that the variations of $g_{+-} \ , , g_{--}$ and $g_{-i}$ all vanish, as required by (\ref{A8}). Consider, as an illustrative example, 
\begin{equation}
\delta g_{--} = -2 g_{\lambda - }\partial_{-}\xi^{\lambda}=
-2 g_{i- }\partial_{-}\xi^{i}=0
\label{A11}
\end{equation}
A nontrivial check may be done for $g_{+i}$ since it is nonvanishing. From (\ref{A9})
\begin{align}
\delta g_{+i} &= -\xi^{\lambda} \partial_{\lambda} g_{+i} - g_{\lambda i} \partial_+ \xi^{\lambda} - g_{\lambda +} \partial_i \xi^{\lambda} \notag\\
&= -\xi^{j} \partial_{j} g_{+i} - g_{j i} \partial_+ \xi^{j}  - g_{j +} \partial_i \xi^{j}
\label{A12}
\end{align} 
Putting the explicit metric components,
\begin{equation}
-\delta\left(\frac{B_i}{m} + \Delta_i\right) = \xi^j \partial_j \left(\frac{B_i}{m} + \Delta_i\right) -g_{ij}\partial_+\xi^j + \left(\frac{B_j}{m} + \Delta_j\right) \partial_i \xi^j
\label{A13}
\end{equation}
Equating terms from both sides yields,
\begin{align}
\delta B_i &= - \xi^j \partial_j B_i - B_j \partial_i \xi^j \notag\\
\delta \Delta_i &= - \xi^j \partial_j \Delta_i + g_{ij}\partial_+ \xi^j - \Delta_j \partial_i \xi^j
\label{A14}
\end{align}
which reproduce the relevant transformations in (\ref{A5}) expressed in light cone coordinates.

Similarly, considering the variation of $g_{++}$, we find from (\ref{A9}),
\begin{align}
\delta g_{++} &= -\xi^{\lambda} \partial_{\lambda} g_{++} - 2 g_{\lambda +} \partial_+ \xi^{\lambda} \notag\\
&= -\xi^{i} \partial_{i} g_{++} - 2 g_{i +} \partial_+ \xi^{i}
\label{A15}
\end{align}
Inserting the explicit structures of the metric components from (\ref{A8}),
\begin{equation}
\delta\left(-\frac{2B_0}{m} + \Delta^i \Delta_i\right) = -\xi^i \partial_i \left(-\frac{2B_0}{m} + \Delta^j \Delta_j\right) + 2 \left(\frac{B_i}{m} + \Delta_i\right) \partial_+ \xi^i\label{A16}
\end{equation}
Equating coefficients yields,
\begin{align}
\delta B_0 &= -\xi^i \partial_i B_0 - B_i \partial_+ \xi^i \notag\\
\delta (\Delta^i \Delta_i)  &= -\xi^i \partial_i(\Delta^j \Delta_j) + 2 \Delta_i \partial_+ \xi^i
\label{A17}
\end{align}
These expressions reproduce the corresponding results given in (\ref{A5}), along with,
\begin{equation}
\delta \Delta^i=\partial_{+}\xi^i-\xi^k\partial_k\Delta^i+\Delta^k
\partial_k\xi^i
\end{equation}
which is obtained from its definition $\Delta_i=g_{ij}\Delta^j$. Finally, the transformation for $\psi$ given in (\ref{A5}) may be reproduced by using the map (23) and the known transformation for the scalar field $\phi$.

We have thus shown that the NRDI action (\ref{A4}) is obtainable as a nonrelativistic limit of the relativistic action (\ref{A1}), exactly as happened for the flat case, once the identification (\ref{A7}) is used. The consistency of such an identification with the transformations (\ref{A5}) was established. Also, observe that once the new fields $(B_0\, ,B_i\, ,\Delta_i)$ introduced by the gauging process are set to zero, the metric (\ref{A8}) reduces to the flat space result and the curved space action (\ref{A4}) goes over to the corresponding flat space theory (24).
\section{Connection with geometry} 
The diffeomorphism invariant non relativistic action (\ref{A4}) differs from the one obtained by minimal coupling  on the spatial slice
\cite{Son:2005rv} in the last term containing $\Delta^i$. Though, the presence of this term has been established from Galilean gauge theory \cite{Banerjee:2014pya,Banerjee:2015rca} and also from a relativistic field theory taking $c\to \infty$ limit \cite{Banerjee:2016bbm},  further justification will always be welcome. Since this term comes from the geometry of spacetime in which the spatial manifold is embedded, one will be naturally interested about the geometry that emerges from the present study. Keeping the discussion of the last section in view, we are just one step behind the task.

We have reduced
a theory (\ref{A2}) in the relativistic Riemannian spacetime to a theory (\ref{A3})using the map (\ref{MAP}). Till then the nature of space time is quite general but our experience in flat spacetime allows us to guess that it is non relativistic in content in one lower dimension. The elements of the metric $g^{\mu\nu}$ is now compared with a Schrodinger field action
coupled with curved space (equation (\ref{A4})). The latter theory is invariant under the transformation (\ref{A10})). If (\ref{A10}))is generalised to include local time translation it becomes,
\begin{equation}
\xi^+ = \xi^+(x^+),{}{}\xi^-=0,{}{}\xi^i = \xi^i(x^+,x^i)
\label{CA1}
\end{equation}
which is the most general form that respects the special role of time in nonrelativistic spacetime . The action
(\ref{A4}) is required to be invariant under (\ref{CA1}) which contains time translation in addition. A new field  $\theta$ has to be introduced to localise it.

 We first rename $\theta = \Sigma_0{}^0$ and $\theta\Delta_i=\Sigma_0{}_i $. Note that $\Sigma_a{}^0 = 0$.The action invariant under(\ref{CA1})is,
\begin{equation}
S= \int dx^0d^3x \det{{\Lambda_\mu{}^\alpha}}\left[\frac{i}{2}(\psi^{*}\nabla_{0}\psi
-\psi\nabla_{{0}}\psi^{*}) -\frac{1}{2m}\nabla_a\psi^{*}
\nabla_a\psi
\right]
\label{localschaction} 
\end{equation}
Here $\nabla_0 ={\Sigma_0}^\mu(\partial_\mu +iB_\mu)$ and $\nabla_a ={\Sigma_a}^k(\partial_k +iB_k)$.
$\Lambda_\mu{}^\alpha$ is the inverse of ${\Sigma_\alpha}^\mu$.

 The invariance of the action (\ref{localschaction})under spacetime diffeomorphism (\ref{CA1}) is derived from the  following transformations of  $\Sigma_\alpha{}^\mu$ and $B_\mu$ \cite{Banerjee:2015rca},
\begin{align}
\delta_0 {\Sigma_0}^{0} &= -\xi^\nu {\partial_\nu{\Sigma}_0}^{0}+ {\Sigma_0}^{\nu}\partial_{\nu}\xi^{0} \nonumber\\
\delta_0 {\Sigma_0}^{k} &= -\xi^\nu {\partial_\nu{\Sigma}_0}^{k}+ {\Sigma_0}^{\nu}\partial_{\nu}\xi^{k} + v^b{\Sigma_b}^{k}\nonumber\\
\delta_0 {\Sigma_a}^{k} &= -\xi^\nu {\partial_\nu{\Sigma}_a}^{k}+ {\Sigma_a}^{\nu}\partial_{\nu}\xi^{k} -\lambda_a{}^b{\Sigma_b}^{k}\nonumber\\
\delta_0 {\Lambda_0}^{0} &= -\xi^\nu {\partial_\nu{\Lambda}_0}^{0}+ {\Lambda_\nu}^{0}\partial_{0}\xi^{\nu}\nonumber\\
\delta_0 {\Lambda_0}^{a} &= -\xi^\nu {\partial_\nu{\Lambda}_0}^{a}+ {\Lambda_\nu}^{a}\partial_{0}\xi^{\nu} - v^a{\Lambda_0}^{0}\nonumber\\
\delta_0 {\Lambda
_k}^{a} &= -\xi^\nu {\partial_\nu{\Lambda}_k}^{a}+ {\Lambda_{\nu}}^a\partial_{k}\xi^{\nu} -\lambda_c{}^a{
\Lambda
_k}^{c}\nonumber\\
\delta_0 {B_k} &= -\xi^\nu \partial_\nu{B_k}- {B_i}\partial_{k}\xi^{i} + m\partial_k(v^ix_i)-mv^b{\lambda_k}^{b}\nonumber\\
\delta_0 {B}_{0} &= -\xi^\nu \partial_\nu B_0 + \partial_{0}\xi^{\mu}B_\mu + m \dot{u}_ix^i+ mv^b{\Lambda_k}^{b}{\Lambda_0}^{0}{\Sigma}_{0}^{k} 
\label{delth3}
\end{align} 
 The new fields $B_\mu$
have the structures,
\begin{eqnarray}
B_\mu = \frac{1}{2}{B_\mu}^{ab}\omega_{ab} + {B_\mu}^{a{0}}\omega_{a}
\label{gaugefields}
\end{eqnarray}
where $\omega_{ab}$ and $\omega_{a}$ are respectively the generators of rotations and Galileo boosts, the corresponding parameters being $\lambda_a{}^b$ and $v^b$, respectively. 
Clearly ${\Sigma_\alpha}^{\mu}$ act as the vierbeins and ${B_\mu}^{ab}$ and ${B_\mu}^{a{0}}$ act as the spin connections.

The geometric interpretation of the action (\ref{A3}) will be obtained from the identification of the inverse metric $g^{\mu\nu}$ and hence the metric. One can then  study  their transformations under (\ref{delth3}) and compare it with (\ref{localschaction}). 
 Our purpose is to show that (\ref{A3}) can be exactly mapped to (\ref{localschaction}). 
under the following map,
\begin{align}
g^{++}&= g^{+i} = 0 \, , \qquad  g^{+ -} = -{\Sigma_0}^{0} \, , \qquad  g^{ij}= g^{ij} \notag\\
g^{--} &= \frac{2}{m}\left({\Sigma_0}^{\mu}B_\mu + \frac{1}{2m}{\Sigma_a}^{\mu}{\Sigma_a}^{\nu}B_\mu B_\nu\right) \,, \quad g^{i-}= -\left(\frac{B^i}{m} + {\Sigma_0}^i\right)
\label{A77}
\end{align}
 if we identify in the nonrelativistic scenario $x^+$ with $x^0$ and $x^i$ denote the spatial coordinates, ($\xi^- =0$, because that dimension is compactified.) then the transformations of the metric elements should follow (\ref{A9}). On the other hand the same transformations should be obtained from the right hand side using (\ref{delth3}). Only if the transformations from both sides match then we can conclude that our goal is achieved.

There are two interesting points. First, putting ${\Sigma_0}^{0}=1 $  we get (\ref{A7}) from (\ref{A77}) and second, in the same limit (\ref{A4} ) is obtained  from (\ref{localschaction}). This shows that the last term of (\ref{A4}) is a necessity, following from the geometry of the space time.

We now look at the mapping (\ref{A77}). It gives us the inverse metric $g^{\mu\nu}$, which is nonsingular. Inversion of  $g^{\mu\nu}$ gives the metric,
\begin{equation}
g_{\mu \nu} = 
\begin{pmatrix}
-\frac{2B_0}{m{\Sigma_0}^0}+\frac{{\Sigma_0}^i{\Sigma_0i} }{{\Sigma_0}^0{}^2} & -1 & -\frac{B_i}{m{{{\Sigma_0}^0}} } -\frac{\Sigma_{0i}}{{{\Sigma_0}^0}}  \\
- 1 & 0 & 0\\
-\frac{B_i}{m{\Sigma_0}^0} -\frac{\Sigma_{0i}}{{{\Sigma_0}^0}}    & 0 & g_{ij}\\
\end{pmatrix}
\label{A88}
\end{equation}
Now observe that we have compactified the $x^-$ dimension and all the fields are $x^-$ independent. So we finally abstract the
space time metric by dropping the row and column corresponding to this $x^-$ coordinate. We can thus propose a single metric for our
nonrelativistic space time,
\begin{equation}
g_{\mu \nu} = 
\begin{pmatrix}
-\frac{2B_0}{m{\Sigma_0}^0}+\frac{{\Sigma_0}^i{\Sigma_0i} }{{\Sigma_0}^0{}^2}  & -\frac{B_i}{m{{{\Sigma_0}^0}} } -\frac{\Sigma_{0i}}{{{\Sigma_0}^0}}  \\
-\frac{B_i}{m{\Sigma_0}^0} -\frac{\Sigma_{0i}}{{{\Sigma_0}^0}}     & g_{ij}\\
\end{pmatrix}
\label{A888}
\end{equation}
 The transformations of the metric tensor may now be worked out using (\ref{delth3}). The new feature is the appearence of the local boost transformations. Remember that when considering NRDI we dropped boost from the local transformations. Due to the presence of local boosts the  transformations of the metric now becomes nontrivial.
 
 We start with $g_{++}$. Its form variation is given by,
 \begin{eqnarray}
 \delta g_{++}& =&-\frac{2}{m}\left[-\frac{1}{{\Sigma_0}^0{}^2}\delta \Sigma_0{}^0 B_0 + \frac{1}{{\Sigma_0}^0 } \delta B_0 \right]- \frac{2}{{\Sigma_0}^0{}^3} \delta {{\Sigma_0}^0}\Sigma_0{}^i \Sigma_{0i}\nonumber\\& +& \frac{1}{{\Sigma_0}^0{}^2}\delta(\Sigma_{0i} {\Sigma_0}^i)\label{CA22}
 \end{eqnarray}
The local boost appears in the variation of $\Sigma_0{}^i$. So the transformation of the last term is crucial. Direct computation gives,
\begin{eqnarray}
\delta(\Sigma_{0i} \Sigma_0{}^i)= \dots\dots + 2v_b\Sigma_{bi} \Sigma_0{}^i)
\label{CA33}
\end{eqnarray}  
The terms which fit in the desired geometric form  are represented by dots on the right hand side of (\ref{CA33}). Using it  in (\ref{CA22})we finally get,
\begin{eqnarray}
\delta g_{++} = -\xi^\nu\partial_\nu g_{++} -2 g_{+\lambda}\partial_+\xi^\lambda + 2v^b\Sigma_{bi} \Sigma_0{}^i
\end{eqnarray}
The last term is clearly an anomalous term. However it can be shown to be equal to $v_0$ which vanishes. Thus local Galilean boost does not disturb the transformation of $g_{++} $.

 This, however, is not true for $g_{+i}$. After some calculations we get,
 \begin{align}
\delta g_{+i} &= -\xi^{\lambda} \partial_{\lambda} g_{+i} -  g_{\lambda i} \partial_+ \xi^{\lambda} \notag\\
&-  g_{+\lambda} \partial_i \xi^{\lambda} + \frac{1}{\Sigma_0{}^0}\partial_i{{(v^k x_k)}}
\label{A155}
\end{align}
The anomalous term 
$\frac{1}{\Sigma_0{}^0}\partial_i (v^k x_k)$
 does not vanish here.

The above analysis shows us that the single non degenerate metric obtained from null reduction is acceptable if Galilean boost is dropped from the set of symmetries. We thus enquire what type of geometry is specified by our metric.

The invariant 'interval' corresponding to the metric (\ref{A888}) helps us to implement a foliation through the Arnowit - Deser - Misner (ADM) construction in general relativity. We define the lapse N and the shift variables in the usual way 
\begin{eqnarray}
N &=&\left(-g^{++}\right)^{1/2}\nonumber\\
N^j&=&g^{ij}g_{+i}
\end{eqnarray}
 Their transformation laws are easily calculated from the above relations,
\begin{align}
\delta g_{ij}&=-\partial_i\xi^k g_{jk}-\partial_j\xi^kg_{ik}-\xi^k
\partial_k g_{ij}-\xi^+{\partial_+ {g}}_{ij},\nonumber\\
\delta N_i&=-\partial_i\xi^jN_j-\xi^j\partial_jN_i-\partial_+\xi^jg_{ij}-\partial_+{\xi}^+ N_i-\xi^+
\partial_+ {N}_i,\nonumber\\
\delta N &=-\xi^j\partial_j N-\partial_+\xi^+ N-\xi^+\partial_+{N}.
\label{foldif}
\end{align}
These are the transformation laws applicable to the projectable Horava gravity \cite{Horava:2009uw}. We thus find that the null reduction leads to a spacetime which acquires boost violating Horava geometry, if we intend to impose a single metric.

\section{Newton-Cartan spacetime}
In the previous section a connection with the geometry of the projectable version of Horava-Lifshitz gravity was established. However there is another nonrelativistic geometry that yields Newton's gravity. This is the Newton-Cartan geometry. The important difference is that whereas Horava-Lifshitz geometry may be obtained from a single nondegenerate metric, the Newton-Cartan geometry requires two degenerate metrics. This is also tied to the fact that, contrary to Newtonian gravity, there is no boost symmetry in the Horava-Lifshitz formalism. Indeed, if all the symmetries are retained, the nonrelativistic case admits two degenerate metrics. This is very simply seen in the flat case where the metric is given by,
\begin{equation}
ds^2=g_{\mu\nu}dx^{\mu}dx^{\nu}=-c^2dt^2+d\overrightarrow{x}^2
\end{equation} 
so that,
\begin{equation}
g_{\mu\nu}=\begin{pmatrix}
-c^2&0&0&0\\0&1&0&0\\0&0&1&0\\0&0&0&1
\end{pmatrix}
\end{equation}
while the inverse metric is given by,
\begin{equation}
g^{\mu\nu}=\begin{pmatrix}
-\frac{1}{c^2}&0&0&0\\0&1&0&0\\0&0&1&0\\0&0&0&1
\end{pmatrix}
\end{equation}
Now taking the nonrelativistic limit $c \rightarrow \infty$ we get a degenerate rank 3 spatial metric from $g^{\mu\nu}$,
\begin{equation}
h^{\mu\nu}=\begin{pmatrix}
0&0&0&0\\0&1&0&0\\0&0&1&0\\0&0&0&1
\end{pmatrix}\label{A}
\end{equation}
and a degenerate temporal `vielbein' of rank 1,
\begin{equation}
\tau_{\mu}=\begin{pmatrix}
-1\\0\\0\\0
\end{pmatrix}\label{P}
\end{equation}
This temporal `vielbein' reproduces the metric $g_{\mu\nu}$ in the limit $c \rightarrow \infty$ as follows,
\begin{equation}
\tau_{\mu\nu}=\tau_{\mu}\tau_{\nu}=\begin{pmatrix}
1&0&0&0\\0&0&0&0\\0&0&0&0\\0&0&0&0
\end{pmatrix}
\end{equation}
which is just $\lim \limits_{c\to \infty}(-\frac{1}{c^2}g_{\mu\nu})$.

Furthermore, the above structures for $h^{\mu\nu}$ and $\tau_{\mu}$ are augmented by $h_{\mu\nu}$ and $\tau^{\mu}$ in the following way,
\begin{equation}
h_{\mu\nu}=\begin{pmatrix}
0&0&0&0\\0&1&0&0\\0&0&1&0\\0&0&0&1
\end{pmatrix}, \quad\tau^{\mu}=(-1,0,0,0)\label{Q}
\end{equation}
It is now possible to verify the relations,
\begin{align}
&h^{\mu\nu}\tau_{\nu}=0,\quad h_{\mu\nu}\tau^{\nu}=0\quad \tau^{\mu}\tau_{\mu}=1\notag\\&h^{\mu\nu}h_{\nu\rho}=
\delta^{\mu}_{\rho}-\tau^{\mu}\tau_{\rho}
\end{align}
The above relations define the algebra of Newton-Cartan geometry. We have provided a flat space representation by taking the NR limit of the flat relativistic metric.

We next show that this representation is obtained by directly taking the flat space limit of the Newton-Cartan metric given by us in the curved background. This metric is contained in the $4\times4$ invertible matrix \cite{Banerjee:2015rca,{Banerjee:2014nja}},
\begin{equation}
{\Sigma_{\alpha}}^{\mu}=\begin{pmatrix}
\theta&\theta\Psi^k\\0&{\Sigma_a}^k
\end{pmatrix}\label{D}
\end{equation}  
The degenerate spatial metric $h^{\mu\nu}$ of rank 3 is given by,
\begin{equation}
h^{\mu\nu}={\Sigma_{a}}^{\mu}{\Sigma_{a}}^{\nu}
\end{equation}
Observing that ${\Sigma_{a}}^{0}=0$ we find that,
\begin{equation}
h^{\mu\nu}=\begin{pmatrix}
0&0\\0&g^{ij}
\end{pmatrix}
\label{lebel1}
\end{equation}
where,
\begin{equation}
g^{ij}={\Sigma_{a}}^{i}{\Sigma_{a}}^{j}
\end{equation}
Here ${\Sigma_{a}}^{i}$ are the vielbein that connect the local coordinates with the global ones. Their product was shown to transform exactly like the spatial metric $g^{ij}$. Taking the flat limit $g^{ij}\rightarrow \delta^{ij}$ shows that (\ref{lebel1}) goes over to (\ref{A}).

The degenerate rank 1 temporal vielbein is given by,
\begin{equation}
\tau_{\mu}=(\frac{1}{\theta}, 0)
\end{equation}
while,
\begin{equation}
\tau^{\mu}={\Sigma_{0}}^{\mu}=(\theta, \theta \Psi^k)
\end{equation}
which are easily identified from (\ref{D}). The fields $\theta$ and $\Psi^{k}$ were introduced to localise the original (global) Galilean symmetry. This was the essence of our method of coupling nonrelativistic theories to gravity since the localised theory admits a geometrical interpretation \cite{Banerjee:2015rca, Banerjee:2014nja}.

The flat space limit is obtained by setting these fields to zero. Then $\lim \limits_{\theta,\psi^k\to 0}(\frac{1}{\theta}\tau^{\mu})=(1,0)$ and $\lim \limits_{\theta\to 0}(\theta\tau^{\mu})=(1,0)$ which reproduce, upto a trivial sign, the corresponding structures in (\ref{P}) and (\ref{Q}).

\section{Conclusions}

The relativistic origin of NR theories has featured quite extensively in recent times \cite{Andreev:2013qsa}-\cite{Ichinose}. While the group theoretical analysis is reasonably well understood, based either on the Inonu-Wigner group contraction or an embediing in one higher dimension using lightcone coordinates, the situation for field theories is much less transparent. In this case also the same two approaches are generally followed. The velocity of light `c' is introduced right from  the beginning in the relativistic model and the limit $c \to \infty$ is taken at the end to abstract the NR theory. Alternatively, the relativistic field theory is formulated in light cone variables and, after an appropriate compactification, the NR theory in one dimension less is deduced. 

It has, however, been noticed that symmetries play a significant role in the reduction process. For instance, it becomes problematic to reduce a real scalar theory to the NR Schroedinger theory since, contrary to the former, the latter has a $U(1)$ gauge invariance. This suggests that a better option would be to start from a complex scalar theory. But as illustrated here, this reduction has its own problems. Presumably, these stem from the fact that a second order (complex scalar) theory is being reduced to a first order (Schroedinger) theory.

The second order theory gets converted into a first order one by invoking light cone variables. Since this change of variables is not a Lorentz transformation, the original relativistic theory gets a `nonrelativistic' look.

We exploited the use of light cone variables to fully analyse the NR reduction of the massless complex scalar theory to the Schroedinger theory. After a physically motivated compactification, we show that there is an exact mapping of the two actions. We then proceed to demonstrate a complete equivalence between the conformal generators of the massless complex scalar theory to the conformal-Galilean generators of the Schroedinger theory. Indeed one may use the present approach to obtain the scale and conformal generators of the Schroedinger theory which, otherwise, require an elaborate calculation \cite{Hagen:1972pd}. 

We have also discussed several other applications of our reduction scheme. The complete Schroedinger symmetry of hydrodynamics was reproduced. The relativistic origin of nonrelativistic diffeomorphism symmetry was elucidated. It was shown that the corresponding geometry was that of the projectable version of Horava-Lifshitz geometry \cite{Horava:2009uw}. A connection with the geometry of Newton's gravity; namely, Newton-Cartan geometry was also shown. This was done by reproducing the flat space NR degenerate metrics from the Newton-Cartan structure found in the NR curved background.


\begin{thebibliography}{999}
\bibitem{Andreev:2013qsa} 
  O.~Andreev, M.~Haack and S.~Hofmann,
  Phys.\ Rev.\ D {\bf 89}, 064012 (2014)
  doi:10.1103/PhysRevD.89.064012
  [arXiv:1309.7231 [hep-th]].


\bibitem{Andreev:2014gia} 
  O.~Andreev,
  Phys.\ Rev.\ D {\bf 91}, no. 2, 024035 (2015)
  doi:10.1103/PhysRevD.91.024035
  [arXiv:1408.7031 [hep-th]].


\bibitem{Jensen:2014wha} 
  K.~Jensen and A.~Karch,
  JHEP {\bf 1504}, 155 (2015)
  doi:10.1007/JHEP04(2015)155
  [arXiv:1412.2738 [hep-th]].


\bibitem{Banerjee:2015rca} 
  R.~Banerjee and P.~Mukherjee,
  Phys.\ Rev.\ D {\bf 93}, no. 8, 085020 (2016)
  doi:10.1103/PhysRevD.93.085020
  [arXiv:1509.05622 [gr-qc]].


\bibitem{Banerjee:2016bbm} 
  R.~Banerjee, S.~Gangopadhyay and P.~Mukherjee,
  Int.\ J.\ Mod.\ Phys.\ A {\bf 32}, no. 19n20, 1750115 (2017)
  doi:10.1142/S0217751X17501159
  [arXiv:1604.08711 [hep-th]].


\bibitem{Horvathy:2009kz} 
  P.~A.~Horvathy and P.-M.~Zhang,
  Eur.\ Phys.\ J.\ C {\bf 65}, 607 (2010)
  doi:10.1140/epjc/s10052-009-1221-x
  [arXiv:0906.3594 [physics.flu-dyn]].


\bibitem{Kaminski:2013gca} 
  M.~Kaminski and S.~Moroz,
  Phys.\ Rev.\ B {\bf 89}, no. 11, 115418 (2014)
  doi:10.1103/PhysRevB.89.115418
  [arXiv:1310.8305 [cond-mat.mes-hall]].


\bibitem{Braaten:2016dlp} 
  E.~Braaten, A.~Mohapatra and H.~Zhang,
  Phys.\ Rev.\ D {\bf 96}, no. 3, 031901 (2017)
  doi:10.1103/PhysRevD.96.031901
  [arXiv:1609.05182 [hep-ph]].


\bibitem{Namjoo:2017nia} 
  M.~H.~Namjoo, A.~H.~Guth and D.~I.~Kaiser,
  ``Relativistic Corrections to Nonrelativistic Effective Field Theories,''
  arXiv:1712.00445 [hep-ph].


\bibitem{Leiva:2003kd} 
  C.~Leiva and M.~S.~Plyushchay,
  Annals Phys.\  {\bf 307}, 372 (2003)
  doi:10.1016/S0003-4916(03)00118-0
  [hep-th/0301244].


\bibitem{Duval:1984cj} 
  C.~Duval, G.~Burdet, H.~P.~Kunzle and M.~Perrin,
  Phys.\ Rev.\ D {\bf 31}, 1841 (1985).
  doi:10.1103/PhysRevD.31.1841

\bibitem{Ichinose}
T.~Ichinose, Jour.\ of Functional Analysis, 73, 233 (1987)
\bibitem{Son:2008ye} 
  D.~T.~Son,
  Phys.\ Rev.\ D {\bf 78}, 046003 (2008)
  doi:10.1103/PhysRevD.78.046003
  [arXiv:0804.3972 [hep-th]].


\bibitem{Horava:2009uw} 
  P.~Horava,
  Phys.\ Rev.\ D {\bf 79}, 084008 (2009)
  doi:10.1103/PhysRevD.79.084008
  [arXiv:0901.3775 [hep-th]].
\bibitem{Hagen:1972pd} 
  C.~R.~Hagen,
  Phys.\ Rev.\ D {\bf 5}, 377 (1972).
  doi:10.1103/PhysRevD.5.377.
   
\bibitem{Duval:2009vt}
C.~Duval and P.~A.~Horvathy,
J. Phys. A42 465206 (2009), doi:10.1088/1751-8113/42/46/465206.
\bibitem{Banerjee:2014pya} 
  R.~Banerjee, A.~Mitra and P.~Mukherjee,
  Phys.\ Lett.\ B {\bf 737}, 369 (2014)
  doi:10.1016/j.physletb.2014.09.004
  [arXiv:1404.4491 [gr-qc]].



\bibitem{Banerjee:2015tga} 
  R.~Banerjee, A.~Mitra and P.~Mukherjee,
  Phys.\ Rev.\ D {\bf 91}, no. 8, 084021 (2015)
  doi:10.1103/PhysRevD.91.084021
  [arXiv:1501.05468 [gr-qc]].



\bibitem{Banerjee:2014nja} 
  R.~Banerjee, A.~Mitra and P.~Mukherjee,
  Class.\ Quant.\ Grav.\  {\bf 32}, no. 4, 045010 (2015)
  doi:10.1088/0264-9381/32/4/045010
  [arXiv:1407.3617 [hep-th]].


\bibitem{Banerjee:2016laq} 
  R.~Banerjee and P.~Mukherjee,
  Class.\ Quant.\ Grav.\  {\bf 33}, no. 22, 225013 (2016)
  doi:10.1088/0264-9381/33/22/225013
  [arXiv:1604.06893 [gr-qc]].


\bibitem{Banerjee:2017rch} 
  R.~Banerjee and P.~Mukherjee,
  ``Milne boost from galilean gauge theory,''
  arXiv:1710.10882 [gr-qc].
  
\bibitem{Son:2005rv} 
  D.~T.~Son and M.~Wingate,
  Annals Phys.\  {\bf 321}, 197 (2006)
  doi:10.1016/j.aop.2005.11.001
  [cond-mat/0509786].
  \end{thebibliography}
\end{document}